\documentclass[twocolumn,showpacs,showkeys,preprintnumbers,amsmath,amssymb]{revtex4}

\usepackage{graphicx}
\usepackage{dcolumn}
\usepackage{bm}
\usepackage{amsmath}
\usepackage{amsfonts}
\usepackage{amssymb}
\usepackage{graphicx}
\usepackage{subfigure}
\usepackage{graphicx}
\usepackage{dcolumn}
\usepackage{bm}

\begin{document}


\title{Resembling holographic dark energy with f(R)gravity as scalar field  and ghost dark energy with tachyon scalar fields  }

\author{ A .Aghamohammadi}
 \email{ a.aqamohamadi@gmail.com }
 \email{  a.aghamohamadi@iausdj.ac.ir}
  \affiliation{ Sanandaj Branch Islamic Azad University,  Iran.}


\date{\today}

\begin{abstract}
In this article,the dynamics and potential of two different scalar  field  models is presented  in a flat Friedmann-Lemaitre-Robertson-Walker(FLRW) universe. One of those models is obtained from the corresponding relation between holographic energy density (HED)  with energy density of the corresponding scalar field of f(R) gravity,and other models  is achieved from connection between the ghost energy  density (GDE) with  energy density of the  tachyon field model. Also, a $f(R)$ model according to the the HDE model is calculated then  stability,anti -gravity and viable conditions on it  are investigated.
\end{abstract}

\maketitle


\section{Introductions}
An approach to investigate the essence of dark energy is well-known to the  HDE density that is proposed by \cite{hsu, ph}  and  have attracted a lot of interest recently, because they relate the dark energy density to the cosmic horizon{\citep{ng, id}. In the quantum field theory $\rho_{\Lambda}$ is regarded as zero-point energy density and defined  based on $L$, the  size of the current universe,(dubbed  the   holographic dark energy )\cite{ph, mi}.
Although there are different choices to IR cutoff such as particle horizon, future horizon, combination of this two case and Hubble horizon we choose the latter, which can satisfy accelerated expansion and alleviate the coincidence $L=H^{-1}$.\\
On other hand, another models of Dark Energy (DE)  has been proposed so- called ghost DE (GDE)
\cite{fr1, agha, cai}. Although,this model is unphysical in the usual Minkowski spacetime,it express the important physical effects in dynamical spacetime. They have shown that in a curved spacetime,the ghost field
gives rise to a vacuum energy density $H{\Lambda}^3$
QCD of the right magnitude $\sim (10^{-3} eV)^4$,
where H is the Hubble parameter and $\Lambda^3$ is QCD mass scale\cite{fr1, fr2, cai1}.\\
 Moreover, it is well known one of the approaches that have been introduced to interpret the present acceleration  to be f(R) theory, which do not seen to introduce new type of matter and can lead to late time acceleration. In fact these theories can be reformulated in terms of scalar tensor theories, with a established coupling of the additional scalar degree of freedom to matter\cite{bi, kha, no, ak, aks, ka, ali, kaa, scn,snd}.
There has been a lot of  attention  in recent years, on establishing a connection between holographic/agegraphic energy density and scalar field models of dark energy \cite{xz, kk, jc, as, as2}. These investigations lead to creation the dynamics and the potentials of the scalar field based on the evolution of holographic and ghost energy density. The author\cite{ash} implement this work, but our work differs from \cite{ash}, in which we  carry out the connection between the HDE and f(R) gravity in the Eneistian  frame which is equivalent with scalar field, also  these similar works  from other point of view have been discussed in the \cite{scn,snd}.\\
In this work we want to study,the corresponding relation between holographic energy density (HED) with energy density of the corresponding scalar field of f(R) gravity by choosing Hubble radius $L=H^{-1}$  as IR cutoff,and connection between the ghost energy  density (GDE) with  energy density of the  tachyon field model. Also, a $f(R)$ model according to the the HDE model is calculated then  stability,anti -gravity and viable conditions on its function  are investigated. Then we able to obtain the clear form of the dynamics of the scalar fields  and  its potentials as a function of time. \\
This paper is organized as follows: At first we review transformation the f(R) gravity models in the equivalent  action of the Jordan conformal frames to the Einstein conformal frame, then we implement a connection between the HDE  and f(R) gravity as scalar field and where is resembled the explicit form of potential $V= V(\phi)$ and the dynamic of the scalar field as a function of time $\phi=\phi(t) $. In the section 3, we make a corresponding relation between the GDE and tachyon field from scalar fields again where is  resembled the clear form of potential, $V=V(t)$  and the dynamic of the scalar field  as function of time $\phi=\phi(t)$ too. Section 4. is devoted to the conclusion.
\section{reconstructing  holographic f(R)gravity model}
 We consider an action of f(R) gravity with general matter  as
\begin{equation}\label{1}
S=\int  \sqrt{-g}\; d^4x\left [\frac{f(R)}{2} +K^2 L_m(\psi, g_{\mu\nu}) \right].
\end{equation}
Where $f(R) $ is an arbitrary function of Ricci scalar, $R$, $L_m = L_m(\psi, g_{\mu\nu})$ is the matter Lagrangian density, $\psi$ is the matter field, $g_{\mu\nu}$ is the metric of space-time, $g$ is the determinant of metric and we have assumed $K^2=8\pi G=1$.
Considering  the equivalence of the f(R) gravity with the scalar field, one can recast this equations in the Einstein frame. In this case, we may use a new set of variables
\begin{eqnarray}\label{g1}
\bar{g}_{\mu\nu}=\Omega g_{\mu\nu},\, \Omega=\frac{df(R)}{dR}=\exp{(-2\beta \phi)}.
\end{eqnarray}
This  conformal map  transforms the above action  in to the Einstein frame{\citep{gm,km}}
\begin{eqnarray}\label{ef3}
S_{EF}&=\frac{1}{2}\int d^4x\sqrt{-g}\Big\{ \bar{R}-\bar{g}_{\mu\nu}\nabla^{\mu}\phi \nabla^{\nu}\phi-2V(\phi)\cr &+2L_m(\bar{g}_{\mu\nu}e^{2\beta\phi}, \psi)\Big\},
\end{eqnarray}
where all indices are raised and lowered by $\bar{g}_{\mu\nu}$. In the Einstein frame, $\phi$ is a minimally coupled scalar field with  a self-interacting potential as
 \begin{equation}\label{4}
V(\phi(R))=\frac{Rf'(R)-f(R) }{2f'^2(R)}.
\end{equation}
In this frame, there exists a coupling between the scalar field, $\phi$ and the matter sector which   its  constant coupling is  $\beta=\sqrt{1/6}$, and being the same for all species of the matter field. A variation with  respect to the metric tensor, $\bar{g}_{\mu\nu}$ gives
\begin{equation}\label{5}
\bar{G}_{\mu\nu}=\bar{T}^{\phi}_{\mu\nu}+\bar{T}^m_{\mu\nu},
\end{equation}
where
\begin{eqnarray}\label{6}
\bar{T}^{\phi}_{\mu\nu}&=&\nabla_{\mu}\phi\nabla_{\nu}\phi-\frac{1}{2}\bar{g}_{\mu\nu}\nabla^{\gamma}\phi\nabla_{\gamma}\phi
-V(\phi)\bar{g}_{\mu\nu}\cr
\bar{T}^m_{\mu\nu}&=&\frac{-2\delta(\sqrt{-\bar{g}}L_m(\bar{g}_{\mu\nu},\psi))}{\sqrt{-\bar{g}}\delta\bar{g}^{\mu\nu}},
\end{eqnarray}
are energy momentum tensor of scalar field and matter field system.It is noticeable that both energy momentum tensor $\bar{T}^{\phi}_{\mu\nu},$ and $\bar{T}^m_{\mu\nu} $ are not separately conserved  namely
\begin{eqnarray}\label{7ef}
\bar{\nabla}^{\mu}\bar{T}^m_{\mu\nu} =-\bar{\nabla}^{\mu}\bar{T}^{\phi}_{\mu\nu}=\beta \bar{T}^m\nabla_{\nu}\phi,
  \end{eqnarray}
but for total energy momentum, i.e, $\bar{T}_{\mu\nu}=\bar{T}^{\phi}_{\mu\nu}+\bar{T}^m_{\mu\nu}$ we have $\bar{\nabla}^{\mu}\bar{T}_{\mu\nu}=0$
 For the flat FLRW universe, the first Friedmann equation of Eq.(\ref{5} ) gives
\begin{eqnarray}\label{8}
3H^2=\rho_{\phi}+\rho_m,
\end{eqnarray}
where $\rho_m$ is  energy density of matter and  $\rho_{\phi}$ is  energy density of a scalar field.\\
 One can define  the energy density and pressure of scalar field  as\cite{snd}
\begin{eqnarray}\label{scf1}
\rho_{\phi}=\frac{1}{2} \dot{\phi}^2+V(\phi) \cr
 p_{\phi}=\frac{1}{2} \dot{\phi}^2-V(\phi).
  \end{eqnarray}
We rewrite Eq. (\ref{7ef}) as
\begin{eqnarray}\label{ef8}
\dot{\rho}_m+3H\rho_m&=&Q \cr
\dot{\rho}_{\phi}+3H\rho_{\phi}(1+\omega_{\phi})&=&-Q,
\end{eqnarray}
where
\begin{eqnarray}\label{scf2}
\omega_{\phi}&=&\frac{p_{\phi}}{\rho_{\phi}}\cr
&=&\frac{3(\dot{R}f''(R))^2-2f'(R)R+2f(R)}{3(\dot{R}f'')^2+2Rf'(R)-2f(R)},
\end{eqnarray}
  and
\begin{equation}\label{9ef}
Q=\beta\dot{\phi}\rho_m.
\end{equation}
It should be noted that unlike other  dark energy models in which the interaction term is manually added to the model, here, there is an external interacting  term between the scalar field and the matter which is presented in Eq.(\ref{9ef})that comes from the model and  is a merit of this scaler field .\\
 To make a correspondence between f(R) gravity in Einstein frame and interacting HDE, we assume the HDE density has the form
\begin{equation}\label{hd1}
\rho_D=3c^2H^2,
\end{equation}
where $c^2$ is a numerical constant that can take the various values  and we have set the system infrared cutoff length, $L$ equal to the Hubble radius
$L=H^{-1}$. Inserting Eq.(\ref{hd1}) in Eq. (\ref{8})gives
\begin{equation}\label{hd2}
r=\frac{1-c^2}{c^2},
\end{equation}
where $r=\rho_m/\rho_D$ in which $\rho _d$ and $\rho _m$ are the dark matter and dark energy density ratios,
respectively. From Eq.(\ref{hd2}),
 Using the time derivative (\ref{hd1}) and (\ref{8}), one can get
\begin{equation}\label{hd4}
  \dot{\rho}_D=-3c^2H\rho_D\Big(1+r+\omega_D \Big)
\end{equation}
By identifying  $\rho_{\phi}=\rho_D$, $\omega_{\phi}=\omega_D$,and combining Eq.(\ref{hd4}) with Eq.(\ref{ef8}), we get
\begin{equation}\label{hd5}
  1+\omega_D=\frac{r(3c^2H-\beta\dot{\phi})}{3H(1-c^2)}.
\end{equation}
On the other hand, from Eqs.(\ref{scf1}, \ref{scf2}) one can get
\begin{equation}\label{scf5}
1+\omega_{\phi}=\frac{\dot{\phi}^2}{\rho_{\phi}},
\end{equation}
and by combining Eqs. (\ref{hd5}, \ref{scf5}) we reach
\begin{equation}\label{om1}
 \dot{\phi}^2=H(3c^2H-\beta\dot{\phi}).
\end{equation}
After some manipulation and integration we obtain
\begin{equation}\label{phi1}
\phi=\eta\ln a,
\end{equation}
where  $\eta=\Big(\sqrt{3c^2+\beta /2}-\beta /2 \Big)$ and we have set $\phi(a_0=1)=0$, that according to (\ref{g1}) we receive    $f(R)=R$.
 Taking the time derivative of (\ref{8}) and using (\ref{hd5}) gives
 \begin{equation}\label{H1}
   \frac{2\dot{H}}{3H^2}=-\frac{\eta^2}{3}+c^2-1,
 \end{equation}
  integrating of (\ref{H1}) gives
 \begin{equation}\label{H2}
  H=\frac{\epsilon}{t}.
 \end{equation}
 Here,
  \begin{equation}\label{epsi}
  \epsilon=2/(\eta^2+3-3c^2)
 \end{equation}


 Integrating again gives
 \begin{equation}\label{H3}
 a=t^{\epsilon}.
 \end{equation}
 Hence Eq. (\ref{phi1}) can be rewritten as
 \begin{equation}\label{H4}
 \phi=\epsilon \eta\ln t.
 \end{equation}
 Now, to obtain the corresponding f(R) function for Eq.(\ref{H4}), we substitute  Eq.(\ref{H2}) and it's time derivative, in to $R=6(\dot{H}+2H^2)$ that gets
 \begin{equation}\label{R1}
   R=\frac{6\epsilon(2\epsilon-1)}{t^2},
 \end{equation}
 Afterwards, by substituting Eq.(\ref{phi1}) in the Eq.(\ref{g1}),  using  Eq.(\ref{R1}) and  integrating we reach
\begin{equation}\label{f2}
f(R)=\frac{1}{(\delta+1){\gamma}{\delta}}R^{\delta+1}\mp\Lambda,
\end{equation}
where $\Lambda$ is the integration constant, $\delta=\epsilon\beta\eta$ and $\gamma=6\epsilon(2\epsilon+1)$.\\
\begin{figure}[h]
\includegraphics[width=0.47\textwidth]{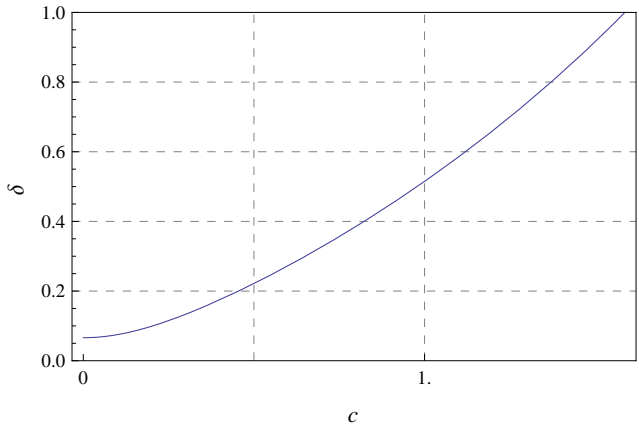}
\caption{\label{fig:F1}The plot shows the evolution of parameter $\delta$ ,Eq.(\ref{epsi}), versus parameter  $c$. The auxiliary parameter is $\beta=\sqrt{1/6}$
.}
\end{figure}
Since the $c$ parameter can take the various values, Figure.1  illustrate the evolution of parameter $\delta$ ,Eq.(\ref{epsi}), versus parameter  $c$.
According to the previous investigates \cite{hos, kha, no4, kaa},the $\delta$ parameter in Eq.(\ref{f2}) should be smaller of one. Hence, Figure .1 shows  for $c\in[0, 1.5]$, we have $\delta\in (0.07- 1)<1$ .\\
 Let us discuss the classical stability condition $d^2f(R)/dR^2=f''(R)$ \cite{1asta} and condition $d f(R)/d R>0$ is required to avoid of anti-gravity regime \citep{2asta, sno1}. Using the $f(R)$ presented in Eq.(\ref{f2}) we have
\begin{equation}\label{f3}
  f'(R)=\frac{1}{{\gamma}^{{\delta}}}R^{\delta}>0,
\end{equation}
It is clear that for $c\in [0-1]$  $\gamma>0$ therefore Eq.(\ref{f3})is always satisfied.
Moreover,for
\begin{equation}\label{f4}
  f''(R)=\frac{\delta}{{\gamma}^{\epsilon\beta\eta}}R^{\epsilon\beta\eta-1}>0,
\end{equation}
As , the $\delta$ and $\gamma$ parameter are  positive constants hence  this condition is satisfied.\\
The several conditions  for the viable f(R) models unifying the late -time acceleration and inflation were proposed in \cite{snd3, snd} that are as.
\begin{enumerate}
\item To enable the current cosmic acceleration  to be generated The current f(R) gravity is considered to be a small constant,\cite{snd3, snd}
\begin{equation}\label{f8}
f(R_0)=2\tilde{R}_0, \quad f'(R_0)\sim 0
\end{equation}
where $R_0\sim(10^{-33}ev)^2$ and $R_0>\tilde{R}_0$ due to the contribution of matter, namely if one  consider $f(R_0)$ as an effective cosmological constant, the effective equation gives $R_0=\tilde{R}_0-k^2 T_{metter}$, here, $ T_{metter}$ is trace of the matter energy- momentum  tensor,  and $f'(R_0)$ not require vanish completely. We take only $|f'(R_0|\ll(10^{-33}ev)^4$

Applying this condition on Eq. (\ref{f2}) give
\begin{equation}\label{f9}
\Big(\delta+1\Big)\ll\frac{10^{-33}R_0}{2\tilde{R}_0\pm \Lambda}.
\end{equation}
\item For existent a flat space -time solution should be satisfied
\begin{equation}\label{f8}
\lim f(R)_{R\rightarrow 0}=0,
\end{equation}
Therefore this constraint restricted our answer in Eq. (ref{f2}) in which$\\Lambda$ has to be zero
\end{enumerate}
At last, the basic consequence of paper is find out the dynamics and potential of scalar field that  using Eq. (\ref{4}) we get
\begin{equation}\label{v}
 V(\phi)=v_1\exp{\Big[\frac{4\delta-2}{\epsilon}\phi}\Big]\pm v_2\exp{[4\beta\eta\phi]},
\end{equation}
where
$$v_1=\gamma^{\delta}\frac{\delta}{1+\delta}\Big[12\epsilon^2-6\epsilon\Big]^{\delta+1},$$
$$v_2=\frac{\Lambda}{12\epsilon(2\epsilon-1)}.$$
Given, \cite{te,vh} the exponential potentials can give rise to an accelerated expansion, hence this potential could lead to an acceleration expansion, and its mass is given by  $m_{\phi^2}=d^2v(\phi)/d{\phi}^2$. One expects the order of the mass $m_{\phi}$ to be that of the Hubble rate, that is, $m_{\phi} \sim H \sim 10^{-33} eV$, which is very light and could make the correction very large  to Newton's law, unless the mass be very large at the local minimum of $v_{\phi}$.

\section{reconstructing ghost tachyon model}
 The tachyon field is one of the dark energy proposals. The equation of state,(EoS) parameter  of a rolling tachyon
stand between $-1$ and $0${\citep{gw}}. Hence , the tachyon field can be appreciated as a reasonable candidate for the
 inflation at high energy {\citep{am}} and be regarded as a source of dark energy depending on the form of the tachyon potential {\citep{tp}}. The correspondence between a tachyon field and various dark energies such  as HDE and ADE has been created{\citep{jc, kk}}, but here we will resemble  the clear form of tachyon potential from GDE  model. As,it has been represented in {\citep{eab}}, the effective lagrangian for the tachyon field is as  follow
 \begin{equation}\label{gt}
L=-V(\phi)\sqrt{1-g^{\mu\nu}\partial_{\mu}\phi \partial_{\nu}\phi},
 \end{equation}
 where $V(\phi)$ is the tachyon potential. The corresponding energy momentum tensor for the tachyon field in the perfect fluid form is given as
 \begin{equation}\label{ta2}
  T_{\mu\nu}=-pg_{\mu\nu}+(\rho+p)u_{\mu}u_{\nu},
 \end{equation}
 where $\rho,\, p$ and  $u_{\mu}$ are the energy density,  pressure  and velocity of the tachyon, respectively, which are given as
 \begin{eqnarray}\label{u1}
   u_{\mu}=\frac{\partial_{\mu}\phi}{\sqrt{\partial_{\nu}\phi\partial^{\nu}\phi}},\cr
   \rho=-T^0_0=\frac{V(\phi)}{\sqrt{1-\dot{\phi}^2}},\cr
   p=T^i_i=-V(\phi)\sqrt{1-\dot{\phi}^2}.
 \end{eqnarray}
From Eq.(\ref{u1}), the EoS parameter of tachyon field is specified as
\begin{equation}\label{wt}
  \omega_t=\frac{p}{\rho}=\dot{\phi}^2-1.
\end{equation}
To obtain the correspondence between GDE and tachyon field, we equate $\omega_D$ with $\omega_t$. S0 by taking the time derivative of $\rho_{\phi}=\rho_D=\alpha H$, where $\alpha$ is a constant of dimension $[energy]^3$ and roughly having order of $\Lambda^3_{QCD}$, in which $\Lambda\sim 100Mev $ is  $QCD$ mass scale, and using Eqs.(\ref{8}, \ref{ef8}), in which $Q=3b^2H\rho_D$, we get
\begin{equation}\label{35}
\omega_D=\frac{3H(2b^2+1)}{\alpha-3H}.
\end{equation}
 By equating Eq.(\ref{wt}) and Eq.(\ref{35}) we achieve
 \begin{eqnarray}\label{36}
\frac{3H(2b^2+1)}{\alpha-3H}=\dot{\phi}^2-1.
 \end{eqnarray}
 To solve Eq.(\ref{36}), we assume $H=k/t$, where $k$ is a constant, so the dynamic of scalar field is given as
 \begin{eqnarray}\label{37}
    \phi(t)&=& \frac{1}{\alpha\sqrt{\alpha t-3k}}\times \\
    &\Big(& (\alpha t-3k)\sqrt{\alpha t +3b^2 k}+\sqrt{\alpha t-3k}(3k+3b^2 k)\cr
    &\log& \Big[\sqrt{\alpha t-3k}+\sqrt{\alpha t+3b^2 k}\Big]\Big)+\phi(t=0),
 \end{eqnarray}
 where $\phi(t=0)$ is the integration constant. Combining Eq.(\ref{37}) and (\ref{u1}) the tachyon potential is specified as
 \begin{equation}\label{38}
   v(t)=\frac{\alpha k}{t}\sqrt{\frac{3k (2b^2+1)}{\alpha t-3k}}.
 \end{equation}
 It is clear that get to  an explicit  expression of $v(t)$, presented in Eq.(\ref{38}),  in terms of $\phi$ is roughly hard, while its time dependant is obtained and we see that the evolution of tachyon is given by $\phi(t)$ and
 $v(t)$.
 Likewise, for $\alpha/3H\ll 1$, it is clear that in Eq. (\ref{36}), we have $\dot{\phi}^2<0$, but in
  the case of $\alpha/3H\gg1$, Eq.(\ref{37}) is reduced as follow
  \begin{eqnarray}\label{39}
     \phi(t) = \phi(t=0)+\sqrt{t+3b^2k}\cr
     \Big[\sqrt{t}+\frac{3b^2k \log [\sqrt{t}+\sqrt{t+3b^2k}]}{\sqrt{t+3b^2k}} \Big],
  \end{eqnarray}
  We see that the evolution of the tachyon and its potential are given by $\phi(t)\propto (t+\log t),\, v(t)\propto 1/\sqrt{t^3-t^2}$.


\section{Conclusion}
In this work by taking the Hubble radius as IR cutoff for interacting HDE model we have obtained a  corresponding relation between the $f(R)$ gravity in the Eniestain frame, as a scalar field model of dark energy, and HDE .Since the $c$ parameter can take the various values. In Fig.1 have been given the interval of the  $c$ parameter values that satisfy the result of the previous investigates \cite{hos, kha, no4, kaa} on the $\delta$ parameter in Eq.(\ref{f2}) should be smaller of one.
At last, our results are summarized as follows
\begin{itemize}
  \item (i)  A analytical form of potential, $V(\phi)$, and dynamics of the  corresponding scalar field for f(R) gravity as an explicit function of time, $\phi(t)$, based on the evolutionary behavior of the interacting HDE model have been resembled,since  the exponential potentials can give rise to an accelerated expansion, hence this potential could lead to an acceleration expansion
  \item Some condition of stability, anti-gravity and viable on the f(R) function obtained to applying some constrain on the $\epsilon $ and $\eta$ parameters have been investigated.
   \item  We have investigated  the corresponding relation between the tachyon scalar field of dark energy and ghost energy density then  the time evolutions of potential, $V(t)$, and dynamics $\phi(t)$ of tachyon field have been obtained. We have seen that due to the complicated form of the $\phi(t)$, it is quite difficult to obtain the analytical form of potential as a function of $\phi$. In addition,  for $\alpha/3H\ll 1$, namely early universe, the $\dot{\phi}^2$ was negative which is impossible, while in the late time that could be $\alpha/3H\gg1$, the dynamics of the scalar field have an explicit form of time that  are given by $\phi(t)\propto (t+\log t),\, v(t)\propto 1/\sqrt{t^3-t^2}$.

      \end{itemize}



\end{document}